\documentclass[twocolumn,showpacs,preprintnumbers]{revtex4}
\usepackage{graphicx}
\usepackage{dcolumn}
\usepackage{bm}
\begin{document}
\newcommand{\ds}{\displaystyle}
\newcommand{\be}{\begin{equation}}
\newcommand{\en}{\end{equation}}
\newcommand{\bea}{\begin{eqnarray}}
\newcommand{\ena}{\end{eqnarray}}
\title{Evolving Lorentzian wormholes supported by phantom matter with constant state parameters}
\author{Mauricio Cataldo}
\altaffiliation{mcataldo@ubiobio.cl} \affiliation{Departamento de
F\'\i sica, Facultad de Ciencias, Universidad del B\'\i o--B\'\i
o, Avenida Collao 1202, Casilla 5-C, Concepci\'on, Chile.}
\author{Sergio del Campo}
\altaffiliation{sdelcamp@ucv.cl} \affiliation{Instituto de F\'\i
sica, Facultad de Ciencias, Pontificia Universidad Cat\'olica de
Valpara\'\i so, Avenida Brasil 2950, Valpara\'\i so, Chile.}
\author{Juan Crisostomo}
\altaffiliation{jcrisostomo@udec.cl} \affiliation{Departamento de
F\'\i sica,
\\ Facultad de Ciencias F\'\i sicas y Matem\'aticas, Universidad
de Concepci\'on, Casilla 160-C, Concepci\'on, Chile.}
\author{Patricio Salgado}
\altaffiliation{pasalgad@udec.cl} \affiliation{Departamento de
F\'\i sica, \\ Facultad de Ciencias F\'\i sicas y Matem\'aticas,
Universidad de Concepci\'on, Casilla 160-C, Concepci\'on, Chile.}
\author{Pedro Labra\~na}
\altaffiliation{plabrana@ubiobio.cl} \affiliation{Departamento de
F\'\i sica, Facultad de Ciencias, Universidad del B\'\i o--B\'\i
o, Avenida Collao 1202, Casilla 5-C, Concepci\'on, Chile.}
\date{\today}
\begin{abstract}
{\bf {Abstract:}} In this paper we study the possibility of
sustaining an evolving wormhole via exotic matter made out of
phantom energy. We show that this exotic source can support the
existence of evolving wormhole spacetimes. Explicitly, a family of
evolving Lorentzian wormholes conformally related to another
family of zero-tidal force static wormhole geometries is found in
Einstein gravity. Contrary to the standard wormhole approach,
where first a convenient geometry is fixed and then the matter
distribution is derived, we follow the conventional approach for
finding solutions in theoretical cosmology. We derive an
analytical evolving wormhole geometry by supposing that the radial
tension (which is negative to the radial pressure) and the
pressure measured in the tangential directions have barotropic
equations of state with constant state parameters. At spatial
infinity this evolving wormhole, supported by this anisotropic
matter, is asymptotically flat, and its slices $t=$ constant are
spaces of constant curvature. During its evolution the shape of
the wormhole expands with constant velocity, i.e without
acceleration or deceleration, since the scale factor has strictly
a linear evolution.

\vspace{0.5cm} \pacs{04.20.Jb, 04.70.Dy,11.10.Kk}
\end{abstract}
\smallskip\
\maketitle \preprint{APS/123-QED}
\section{Introduction}
Wormholes, as well as black holes, are an extraordinary
consequence of Einstein's equations of general relativity. During
recent last decades, there has been a considerable interest in the
field of wormhole physics. Two separate directions emerged: one
relating to Euclidean signature metrics~\cite{Coleman,Strominger}
and the other concerned with Lorentzian ones. The interest has
been focused on traversable Lorentzian wormholes (which have no
horizons, allowing two-way passage through them), and were
especially stimulated by the pioneering works of Morris, Thorne
and Yurtsever~\cite{Morris}, where static, spherically symmetric
Lorentzian wormholes were defined and considered to be an exciting
possibility for constructing time machine models with these exotic
objects, for backward time travel (see also~\cite{Novikov}).

Most of the efforts are directed to study static configurations
that must have a number of specific properties in order to be
traversable. The most striking of these properties is the
violation of energy conditions. This implies that the matter
supporting the traversable wormholes is
exotic~\cite{Morris,Visser}, which means that it has very strong
negative pressures, or even that the energy density is negative,
as seen by static observers. However, one can also consider
time-dependent wormhole configurations, such as rotating
wormholes~\cite{Teo} or evolving wormholes in a cosmological
background~\cite{Kar1,Kar2,Lobo,Arellano}.

Lower~\cite{Kim} and higher dimensional wormholes have also been
considered by several authors. Euclidean wormholes have been
studied by Gonzales--Diaz and by Jianjun and
Sicong~\cite{Gonzales} for example. The Lorentzian ones have been
studied in the context of the n--dimensional Einstein
theory~\cite{Cataldo} or Einstein--Gauss--Bonnet theory of
gravitation~\cite{Biplab}. Evolving higher dimensional wormholes
also have been studied~\cite{Sayan}.

The theoretical construction of wormhole geometries is usually
performed by using the method where, in order to have a desired
metric, one is free to fix the form of the metric functions, such
as the redshift and shape functions, or even the scale factor for
evolving wormholes. In this way one may have a redshift function
without horizons, or with a desired asymptotic. Unfortunately, in
this case we can obtain expressions for the energy and pressure
densities which are physically unreasonable.

In this paper we shall follow the conventional method for finding
solutions in general relativity, and used also in theoretical
cosmology. We shall prescribe the matter content by specifying the
equations of state of the radial and the tangential pressures and
then we solve the Einstein field equations in order to find the
redshift and shape functions together with the scale factor.
Specifically we shall consider that these pressures obey
barotropic equations of state with constant state parameters. In
other words, we shall find all evolving wormhole geometries which
have the radial and the tangential pressures proportional to the
energy density.

The outline of the present paper is as follows: In Sec. II we
briefly review some important aspects of static wormholes and give
the definition of evolving wormholes. In Sec. III we find the
metric of evolving wormholes with pressures obeying barotropic
equations of state with constant state parameters. In Sec. IV the
properties of the obtained wormhole geometry are studied. We use
the metric signature ($-+++$) and set $c=1$.

\section{Evolving Lorentzian wormholes}
\subsection{Characterization of a static Lorentzian wormhole}
Before treating evolving Lorentzian wormholes let us review the
static ones. The metric ansatz of Morris and Thorne~\cite{Morris}
for the spacetime which describes a static Lorentzian wormhole is
given by
\begin{eqnarray}\label{4wormhole}
ds^2=-e^{2\Phi(r)}dt^2+\frac{dr^2}{1-\frac{b(r)}{r}}+r^2(d\theta^2+sin^2
\theta d \varphi^2),
\end{eqnarray}
where $\Phi(r)$ is the redshift function, and $b(r)$ is the shape
function since it controls the shape of the wormhole.

Morris and Thorne have discussed in detail the general constraints
on the functions $b(r)$ and $\Phi(r)$ which make a
wormhole~\cite{Morris}:

Constraint 1: A no--horizon condition, i.e. $e^{\Phi(r)}$ is
finite throughout the space--time in order to ensure the absence
of horizons and singularities.

Constraint 2: The shape function $b(r)$ must obey at the throat $r
= r_0$ the following condition: $b(r_0) = r_0$, being $r_0$ the
minimum value of the $r$--coordinate. In other words
$g^{-1}_{rr}(r_0)=0$.

Constraint 3: Finiteness of the proper radial distance, i.e.
\begin{eqnarray}\label{r finito}
\frac{b(r)}{r} \leq 1,
\end{eqnarray}
(for $r \geq r_0$) throughout the space--time. This is required in
order to ensure the finiteness of the proper radial distance
$l(r)$ defined by
\begin{eqnarray}
l(r)=\pm \int^r_{r_0} \frac{dr}{\sqrt{1-b(r)/r}}.
\end{eqnarray}
The $\pm$ signs refer to the two asymptotically flat regions which
are connected by the wormhole. The equality sign in~(\ref{r
finito}) holds only at the throat.

Constraint 4: Asymptotic flatness condition, i.e. as $l
\rightarrow \pm \infty$ (or equivalently, $r \rightarrow \infty$)
then $b(r)/r \rightarrow 0$.

Notice that these constraints provide a minimum set of conditions
which lead, through an analysis of the embedding of the spacelike
slice of~(\ref{4wormhole}) in a Euclidean space, to a geometry
featuring two asymptotically flat regions connected by a bridge.

Although asymptotically flat wormhole geometries have been
extensively considered in the literature, one can study however
other asymptotic behaviors that are worth considering. For
instance asymptotically anti--de Sitter wormholes may be also of
particular interest~\cite{Barcelo}.


\subsection{Evolving Lorentzian wormholes}
We shall consider a simple generalization of the original Morris
and Thorne metric~(\ref{4wormhole}) to a time-dependent metric
given by
\begin{eqnarray}\label{evolving wormhole}
ds^2= \nonumber\\  -e^{2\Phi(r)}dt^2+a(t)^2 \left(
\frac{dr^2}{1-\frac{b(r)}{r}}+r^2(d\theta^2+sin^2 \theta d
\varphi^2)\right),
\end{eqnarray}
where $a(t)$ is the scale factor of the universe. Note that the
essential characteristics of a wormhole geometry are still encoded
in the spacelike section. It is clear that if $b(r) \rightarrow 0$
and $\Phi(r) \rightarrow 0$ the metric~(\ref{evolving wormhole})
becomes the flat Friedmann-Robertson-Walker (FRW) metric, and as
$a(t) \rightarrow const$ it becomes the static wormhole
metric~(\ref{4wormhole}).

In general, in order to construct an evolving wormhole, one has to
specify or determine the red--shift function $\Phi(r)$, the shape
function $b(r)$ and the scale factor $a(t)$. So, one of them may
be chosen by fiat and the others may be determined by implementing
some physical conditions. For example in Ref.~\cite{Roman15} an
exponential scale factor is considered in order to explore the
possibility that inflation might provide a natural mechanism for
the enlargement of an initially small (possibly submicroscopic)
wormhole to macroscopic size. In Ref.~\cite{Kar1} also different
choices for the scale factor $a(t)$ are considered and the
constraints are found on the minimum values of the throat radii.

In this paper we shall require that $\Phi(r)=0$ in order to have a
family of evolving Lorentzian wormholes conformally related to
another family of zero--tidal force static wormholes, and to
ensure that there is no horizon. We also shall require that the
radial tension, which is the negative of the radial pressure, and
the pressure measured in the tangential directions (orthogonal to
the radial direction) have barotropic equations of state with
constant state parameters. These simple choices will permit us to
find explicit analytical expressions, by solving the Einstein
field equations, for the shift and shape functions, the scale
factor, and the energy and pressure densities.

\section{Einstein field equations for the evolving Lorentzian wormholes}
In order to simplify the analysis and the physical interpretation
(with $\Phi(r)=0$) we now introduce the proper orthonormal basis
as
\begin{eqnarray}
ds^2= - \theta^{(t)}
\theta^{(t)}+\theta^{(r)}\theta^{(r)}+\theta^{(\theta)}\theta^{(\theta)}+
\theta^{(\varphi)}\theta^{(\varphi)},
\end{eqnarray}
where the basis one--forms $\theta^{(\alpha)}$ are given by
\begin{eqnarray}
\theta^{(t)}= dt; \,\,\,\,\, \theta^{(r)}=
\frac{a(t) \, dr}{\sqrt{1-\frac{b(r)}{r}}}; \nonumber \\
\theta^{(\theta)}=a(t) \, r d \theta;  \,\,\,\,\,
\theta^{(\varphi)}= a(t) \, r \, sin \, \theta \,d \varphi.
\end{eqnarray}
These basis one--forms are related to the following set of
orthonormal basis vectors defined by
\begin{eqnarray}\label{basis quieta}
e_{\hat{t}}= e_{t}; \,\,\,\,\, e_{\hat{r}}= a(t)^{-1}
\sqrt{1-\frac{b(r)}{r}} \, e_{r}; \nonumber \\
e_{\hat{\theta}}= a(t)^{-1} r^{-1} \, e_{\theta};
e_{\hat{\varphi}}= a(t)^{-1} \, r^{-1} \, sin^{-1}\theta \,
e_{\varphi}.
\end{eqnarray}
This basis represents the proper reference frame of a set of
observers who always remain at rest at constant $r$, $\theta$,
$\varphi$~\cite{Roman15}.

For these basises the only nonzero components of the
energy--momentum tensor $T_{(\mu)(\nu)}$ are precisely the
diagonal terms $T_{(t)(t)}$, $T_{(r)(r)}$, $T_{(\theta)(\theta)}$
and $T_{(\varphi)(\varphi)}$, which are given by
\begin{eqnarray}\label{EMT}
T_{(t)(t)}=\rho(t,r),
T_{(r)(r)}=p_r(t,r)=-\tau(t,r),\nonumber \\
T_{(\theta)(\theta)}=T_ {(\varphi)(\varphi)}=p_{_l}(t,r),
\end{eqnarray}
where the quantities $\rho(t,r)$, $p_r(t,r)$,
$\tau(t,r)(=-p_r(t,r))$, and
$p_{_l}(t,r)(=p_{\varphi}(t,r)=p_{\theta}(t,r))$ are respectively
the energy density, the radial pressure, the radial tension per
unit area, and lateral pressure as measured by observers who
always remain at rest at constant $r$, $\theta$, $\varphi$.

Thus for the spherically symmetric wormhole metric~(\ref{evolving
wormhole}), with $\Phi(r)=0$, the Einstein equations are given by
\begin{eqnarray} \label{00}
\kappa \rho(t,r)= 3 H^2+\frac{b^{\prime}}{a^2 r^2}, \\
\kappa p_r(t,r)=- \kappa \tau(t,r)= -2 \, \frac{\ddot{a}}{a} -H^2-\frac{b}{a^2 r^3}, \label{rr}\\
\kappa p_{_l}(t,r)= -2 \, \frac{\ddot{a}}{a} -H^2+\frac{b-r
b^{\prime}} {2 a^2 r^3},\label{thetatheta}
\end{eqnarray}
where $\kappa=8 \pi G$, $H=\dot{a}/a$, and an overdot and a prime
denote differentiation with respect to $t$ and $r$ respectively.

Now we shall require that the radial tension and the lateral
pressure have barotropic equations of state. Thus we can write
\begin{eqnarray} \label{C00}
\tau(t,r)=-p_r(t,r)=-\omega_r \, \rho(t,r), \nonumber \\
p_{_l}(t,r)=\omega_{_l} \, \rho(t,r),
\end{eqnarray}
where $\omega_r$ and $\omega_{_l}$ are constant state parameters.
Clearly, the requirement~(\ref{C00}) with $\omega_r=\omega_{_l}$
allows us to connect the evolving wormhole
spacetime~(\ref{evolving wormhole}) with the standard FRW
cosmologies, where the isotropic pressure density is expressed as
$p=\omega \rho$, with constant state parameter $\omega$
(=$\omega_r=\omega_{_l}$).

Now, using the conservation equation $T^\mu_{\,\,\,\,\nu;\mu}=0$,
we have that
\begin{eqnarray}\label{CE1}
\dot{\rho}+H (3 \rho+p_r+2p_{_l})=0, \\
\frac{2(p_{_l}-p_r)}{r}=\frac{2(p_{_l}+\tau)}{r}=p^{\, \prime}_r,
\label{CE2}
\end{eqnarray}
which may be interpreted as the conservation equation and the
relativistic Euler equation (or the hydrostatic equation for
equilibrium for the matter supporting the wormhole) respectively.
From these equations we see that for
$\omega_r=\omega_{_l}=\omega$, i.e. $p_{_l}=p_r=p$, we have the
standard cosmological conservation equation $\dot{\rho}+3 H
(\rho+p)=0$, with $p^{\, \prime}_r=0$, so if we want to isotropize
the pressure with a barotropic equation of state and constant
state parameters, then we can not have a pressure of the form
$p=p(t,r)$, it will depend only on time $t$.

Now, with the help of the conservation equation and the
relativistic Euler equation we can easily solve the Einstein
equations~(\ref{00})--(\ref{thetatheta}). From the structure of
these conservation equations we see that one can write the energy
density in the form $\rho(t,r)=\rho_t(t)\rho_r(r)$. Thus from the
conservation equation we obtain
\begin{equation}\label{rhot}
\rho_t(t)=C_1 a^{-(3+\omega_r+2\omega_{_l})},
\end{equation}
where $C_1$ is an integration constant. Now taking into account
Eq.~(\ref{C00}), from Eq.~(\ref{CE2}) we have that
\begin{equation}\label{rhor}
\rho_r(r)=C_2 r^{2(\omega_{_l}-\omega_r)/\omega_r},
\end{equation}
where $C_2$ is an integration constant. Thus from
expressions~(\ref{rhot}) and~(\ref{rhor}) we can write for the
energy density
\begin{equation}\label{rho}
\rho(t,r)=C \, r^{2(\omega_{_l}-\omega_r)/\omega_r} \,
a^{-(3+\omega_r+2\omega_{_l})},
\end{equation}
where we have introduced a new constant $C$ in order to redefine
the integration constants $C_1$ and $C_2$.

Now, by subtracting Eqs.~(\ref{rr}) and~(\ref{thetatheta}), and
using Eq.~(\ref{00}), we obtain the differential equation
\begin{equation}
\frac{\kappa (\omega_{_l}-\omega_r)\, C \,
r^{2(\omega_{_l}-\omega_r)/\omega_r}}{a^{(3+\omega_r+2\omega_{_l})}}=\frac{3b-r
b^{\prime}}{2a^2r^3}.
\end{equation}
Clearly, from this equation we conclude that if we want to have a
solution for the shape function $b=b(r)$ we must constrain the
state parameters $\omega_r$ and $\omega_{_l}$ in the following
manner:
\begin{equation}\label{constraint}
\omega_r+2\omega_{_l}+1=0,
\end{equation}
thus obtaining for the shape function
\begin{equation}\label{br}
b(r)=Dr^3-\kappa \, C \, \omega_r \, r^{-1/\omega_r},
\end{equation}
where $D$ is a new integration constant.

Now, from Eqs.~(\ref{00}),~(\ref{rho}),~(\ref{br}) and taking into
account the constraint~(\ref{constraint}) we find that the scale
factor is given by
\begin{equation}\label{scf}
a(t)=\sqrt{-D}t+F,
\end{equation}
where $F$ is an integration constant, obtaining the following
final expression for the energy density~(\ref{rho}):
\begin{equation}\label{rhofinal}
\rho(t,r)=\frac{C \,
r^{-(1+3\omega_r)/\omega_r}}{(\sqrt{-D}t+F)^{2}}.
\end{equation}
Notice that in principle one would expect the scale factor to have
the form $a(t)=Et+F$, where $E$ is a constant, but the field
equations constrain this constant to be $E=\sqrt{-D}$.

Thus the self--consistent solution for constant state parameters
$\omega_r$ and $\omega_{_l}$ is given by
Eqs.~(\ref{scf}),~(\ref{br}) and~(\ref{rhofinal}), so obtaining
for the line element~(\ref{evolving wormhole}) the following
wormhole metric:
\begin{eqnarray}\label{evolving wormhole final}
ds^2=-dt^2+(\sqrt{-D}t+F)^2 \times \nonumber\\ \left(
\frac{dr^2}{1+\kappa C \omega_r r^{-(1+\omega_r)/\omega_r}-D
r^2}+r^2(d\theta^2+sin^2 \theta d \varphi^2)\right).
\end{eqnarray}
In this case the constraint~(\ref{constraint}) implies that the
radial and tangential pressures are given by
\begin{equation}\label{pr pl}
p_r=\omega_r \rho, p_{_l}= -\frac{1}{2} \, (1+\omega_r) \rho,
\end{equation}
so the energy density and pressures satisfy the following
relation:
\begin{equation}
\rho+p_r+2 p_{_l}=0.
\end{equation}
Note that there is another branch of spherically symmetric
solutions to Eqs.~(\ref{00})--(\ref{thetatheta}). By adding these
equations and taking into account Eqs.~(\ref{C00}) and~(\ref{rho})
we obtain the equation
\begin{equation}\label{acceleration}
6\frac{\ddot{a}}{a}=-\kappa (1+\omega_r+2\omega_{_l})\,C \,
r^{2(\omega_{_l}-\omega_r)/\omega_r} \,
a^{-(3+\omega_r+2\omega_{_l})},
\end{equation}
which implies that we must take $\omega_r=\omega_{_l}=\omega$,
thus obtaining from Eq.~(\ref{rho}) that $\rho=C \,
a^{-3(1+\omega)}$ and, for the scale factor
$a(t)=(At+B)^{2/(3(1+\omega))}$, i.e. the standard FRW solution
for an ideal fluid with $p(t)=\omega \rho(t)$.

\section{Wormhole solutions}
One interesting aspect to be considered is the possibility of
sustaining a traversable wormhole in spacetime via exotic matter
made out of phantom energy. The latter is considered as a possible
candidate for explaining the late time accelerated expansion of
the Universe~\cite{Cataldo15}. This phantom energy has a very
strong negative pressure and violates the null energy condition,
so becoming a most promising ingredient to sustain traversable
wormholes.

Notice however that in this case we shall use the notion of the
phantom energy in a more extended sense since, strictly speaking,
the phantom matter is a homogeneously distributed fluid, and here
it will be an inhomogeneous and anisotropic
fluid~\cite{Sushkov,Lobo1}, since $p_{r}<-1$, and $p_{_l} \neq
p_r$.

Now we shall discuss the above obtained analytical solution. To
start with, we shall consider first the static case.

\subsection{Static wormhole geometries}
It is clear that for $D=0$ (without any loss of generality we can
set $F=1$) we have a static spherically symmetric spacetime. From
the condition for the throat that the $r$--coordinate has a
minimum at $r_0$, i.e. $g^{-1}_{rr}(r_0)=0$, we obtain for the
integration constant $C=-\frac{r_0^{(1+\omega_r)/\omega_r}}{\kappa
\omega_r}$, yielding for the shape function and the energy density
\begin{eqnarray}\label{whstatic}
b(r)=r_0\left(\frac{r}{r_0}\right)^{-1/\omega_r},
\kappa \rho(r)=-\frac{(r/r_0)^{-(1+3\omega_r)/\omega_r}}{r^2_0
\omega_r},
\end{eqnarray}
respectively. In this case the metric is given by
\begin{eqnarray}\label{evolving wormhole final estatico}
ds^2=-dt^2+ \left(
\frac{dr^2}{1-(r/r_0)^{-(1+\omega_r)/\omega_r}}+ \right. \nonumber \\
\left. r^2(d\theta^2+sin^2 \theta d \varphi^2) \frac{}{}\right).
\end{eqnarray}
The radial coordinate $r$ has a range that increases from a
minimum value at $r_0$, corresponding to the wormhole throat, to
infinity. From Eqs.~(\ref{whstatic}) and~(\ref{evolving wormhole
final estatico}) we can see that for a matter content with a
radial pressure having a phantom equation of state, i.e. $\omega_r
< -1$, we have an asymptotically flat wormhole with a positive
energy density. This static wormhole solution is a traversable one
and was firstly considered in Ref.~\cite{Lobo1}. For $\omega_r>0$
we also have an asymptotically flat wormhole spacetime, but in
this case the energy density is negative everywhere.

\subsection{Evolving wormhole geometries}
Let us now explore the features of the evolving wormhole. We shall
consider the time interval $0<t<\infty$ for the evolution. In
order to maintain the Lorentzian signature we must require that $D
\leq 0$; if $D \geq 0$ the signature of the spacetime changes to a
Euclidean one, obtaining an evolving Euclidean wormhole.

Clearly, in order to have an evolving wormhole, as in the static
case, we must require $\omega_r < -1$ or $\omega_r >1$, yielding
in both these cases that $(1+\omega_r)/\omega_r>0$. Thus we
conclude that the phantom energy can support the existence of
evolving wormholes.

Now it can be shown that for $D < 0$ and $C \omega_r < 0$ the
metric component $g^{-1}_{rr}=1+\kappa C \omega_r
r^{-(1+\omega_r)/\omega_r}-D r^2$ of the line
element~(\ref{evolving wormhole final}) is equal to zero for some
value of the radial coordinate. Effectively, from the formulated
above constraints on the parameters, i.e. $\omega_r<-1$, $C>0$ and
$D<0$, we have that $g^{-1}_{rr}<0$ at the vicinity of $r \gtrsim
0$, while its first derivative $d g^{-1}_{rr}/dr
>0$. This implies that for any $r > 0$ we have always a growing
$g^{-1}_{rr}$. Thus we conclude that for some $0<r_0< \infty$ we
have $g^{-1}_{rr}(r_0)=0$, implying that at the location $r=r_0$
is the throat of the wormhole. So, from the condition
$g^{-1}_{rr}(r=r_0)=0$, we obtain for the integration constant
\begin{eqnarray}
C=\frac{(Dr^2_0-1)}{\kappa \omega_r} \,
r_0^{(1+\omega_r)/\omega_r},
\end{eqnarray}
yielding for the shape function, the metric component $g_{rr}$ and
the energy density
\begin{eqnarray}\label{whevolving}
b(r)=r_0\left(\frac{r}{r_0}\right)^{-1/\omega_r}
\,\,\,\,\,\,\,\,\,\,\,\,\,\,\,\,\,\,\,\,\,\,\,\,\,\,\,\,\,\,
\nonumber
\\ +Dr^3_0 \left(\frac{r}{r_0}\right)^3
\left(1-\left(\frac{r}{r_0}\right)^{-(1+3\omega_r)/\omega_r}\right),
\nonumber \\ g_{rr}^{-1}= 1-
\left(\frac{r}{r_0}\right)^{-(1+\omega_r)/\omega_r}
\,\,\,\,\,\,\,\,\,\,\,\,\,\,\,\,\,\,\,\,\,\,\,\,\,\,\,\,\,\,
\nonumber
\\ -Dr^2_0 \left(\frac{r}{r_0}\right)^2
\left(1-\left(\frac{r}{r_0}\right)^{-(1+3\omega_r)/\omega_r}\right),
\nonumber \\  \kappa \rho(t,r)=\frac{1-Dr^2_0}{\omega_r r^2_0
(\sqrt{-D}t^2+F)^2}
\left(\frac{r}{r_0}\right)^{-(1+3\omega_r)/\omega_r},
\end{eqnarray}
respectively.

Let us now enumerate some characteristic properties of the found
evolving wormhole geometry:

(i) The weak energy condition (WEC) for the energy--momentum
tensor~(\ref{EMT}) reduces to the following inequalities
\begin{eqnarray}
\rho(t,r)\geq0, \,\,\, \rho(t,r)+p_r(t,r)\geq0, \nonumber \\
\rho(t,r)+p_{_l}(t,r)\geq0,
\end{eqnarray}
for all (t,r). By using the expressions~(\ref{C00})
and~(\ref{constraint}) we can rewrite the WEC as follows
\begin{eqnarray}\label{WEC}
\rho(t,r)\geq0, \,\,\, (1+\omega_r) \, \rho(t,r) \geq0, \nonumber \\
(1-\omega_r) \, \rho(t,r)  \geq0.
\end{eqnarray}
Thus for $\omega_r<-1$ the first and third inequalities
of~(\ref{WEC}) are satisfied, while the second one is violated.
So, as one would expect, these evolving wormholes, supported by an
anisotropic phantom energy, do not avoid the violation of the WEC.

(ii) The general form of the evolving wormhole solution implies
that there is only the standard coordinate singularity at the
throat, although for any $t=$ const, the radial proper length
between any two points $r_1$ and $r_2$
\begin{eqnarray}
l(t) = \pm \, a(t)  \int_{r_1}^{r_2} \frac{dr}{\sqrt{1-Dr^2+\kappa
C \omega_r r^{-(1+\omega_r)/\omega_r} }},
\end{eqnarray}
with $r_1\geq r_0$, is required to be finite everywhere. There
are, however, no spatial and temporal curvature singularities
($F>0$). The energy density also is well behaved since at
$(t,r)=(0,r_0)$ it is given by $\rho=C F^{-2}\,
r_0^{-(1+3\omega_r)/\omega_r}$. A temporal singularity occurs at
$t=0$ only for the case with $F=0$.

(iii) From Eq.~(\ref{acceleration}) and the
constraint~(\ref{constraint}) we conclude that the expansion of
the wormhole is not accelerated. So this family of evolving
wormholes, supported by an anisotropic phantom energy, expands
with a constant velocity. Note that from Eq.~(\ref{pr pl}) we have
that if $\omega_r<-1$ then always $p_{_l} >0$, while $p_r <0$.

(iv) From the metric~(\ref{evolving wormhole final}) we can see
that for wormholes supported by phantom matter at spatial infinity
($r\rightarrow \infty$) we have the following asymptotic metric:
\begin{eqnarray}\label{wormhole asintotico}
ds^2 \approx -dt^2+ \nonumber\\ (\sqrt{-D}t+ F)^2  \left(
\frac{dr^2}{1-D r^2}+r^2(d\theta^2+sin^2 \theta d
\varphi^2)\right).
\end{eqnarray}
This metric has slices $t=$ const which are spaces of constant
curvature. This implies that the asymptotic metric~(\ref{wormhole
asintotico}) is foliated with spaces of constant curvature. So the
form of the $r$--dependent part of this metric may induce us to
think that we have a four dimensional spacetime of constant
curvature, implying that we have not an asymptotically flat
wormhole. Namely, since we have that $D<0$, we would have an
asymptotically anti--de Sitter spacetime.

However, if we calculate the Riemann tensor for the
metric~(\ref{evolving wormhole final}) we find that its
independent non--vanishing components are
\begin{eqnarray}\label{CRT}
R_{(\theta)(\varphi)(\theta)(\varphi)}=\frac{\kappa C \omega_r \,
r^{-(1+3\omega_r)/\omega_r}}{(\sqrt{-D}t+F)^2}, \nonumber \\
R_{(r)(\varphi)(r)(\varphi)}=R_{(r)(\theta)(r)(\theta)}= \nonumber \\
-\frac{\kappa C (1+\omega_r) \,
r^{-(1+3\omega_r)/\omega_r}}{2(\sqrt{-D}t+F)^2}.
\end{eqnarray}
From these expressions we see that at spatial infinity these
components vanish for a wormhole supported by a phantom matter.
Since the energy density~(\ref{rhofinal}) also vanishes for $r
\rightarrow \infty$ thus we have an asymptotically flat evolving
wormhole. Notice that we obtain such an asymptotic behavior since
the integration constant $D$ in Eq.~(\ref{br}) finally is
constrained by the field equations to appear also in the general
expression for the scale factor~(\ref{scf}). Thus the asymptotic
metric~(\ref{wormhole asintotico}) can be carried explicitly to
the Minkowski--form metric
\begin{eqnarray*}
ds^2=-d\tau^2 + d\rho^2+\rho^2(d\theta^2+sin^2 \theta d \varphi^2)
\end{eqnarray*}
with the help of the transformation
\begin{eqnarray}
t=\sqrt{\tau^2-\rho^2}-\frac{F}{\sqrt{-D}}, \nonumber \\
r=\frac{\rho}{\sqrt{-D(\tau^2-\rho^2)}}.
\end{eqnarray}

\begin{figure}
\includegraphics[width=3.15in]{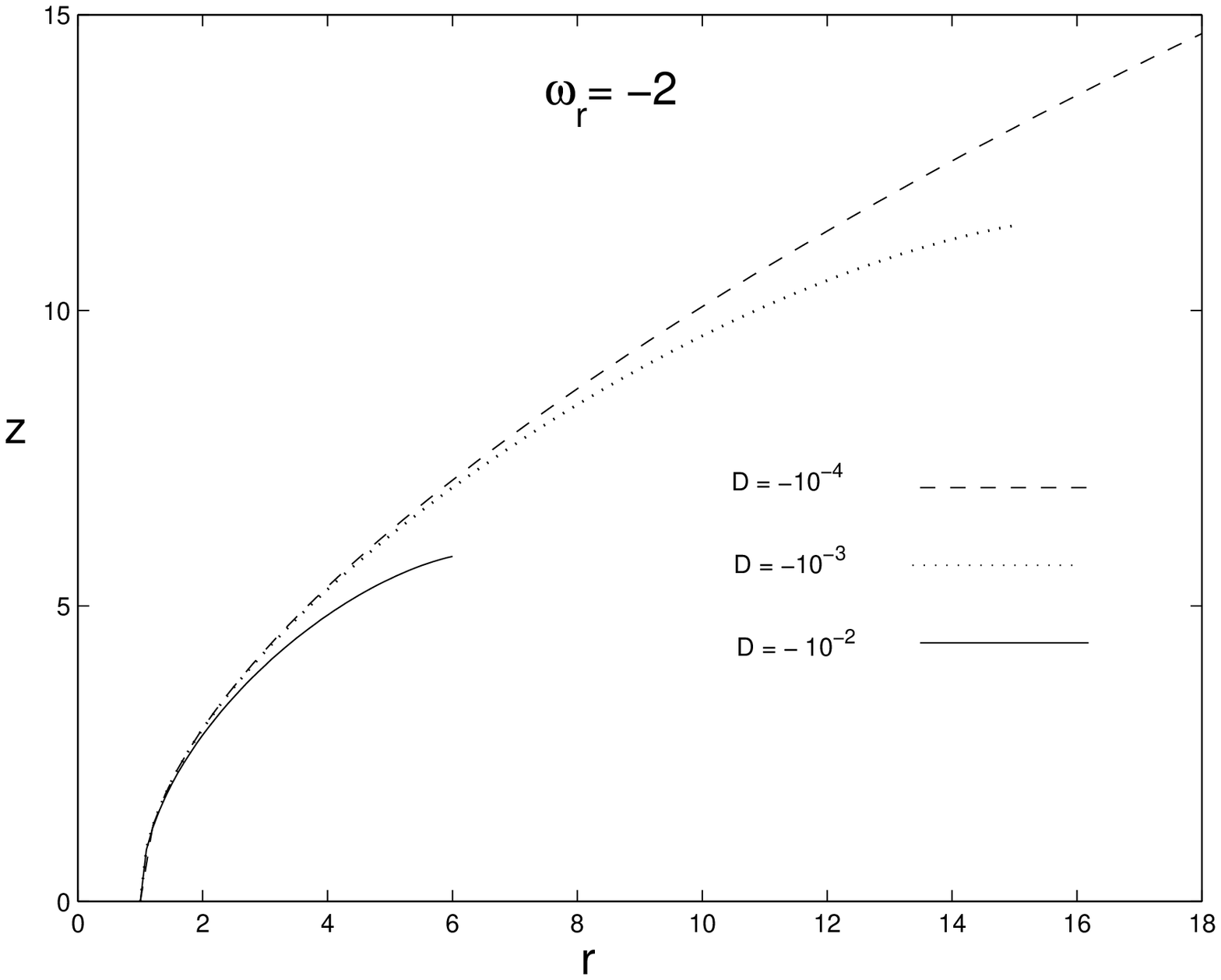} \\
\includegraphics[width=3.15in]{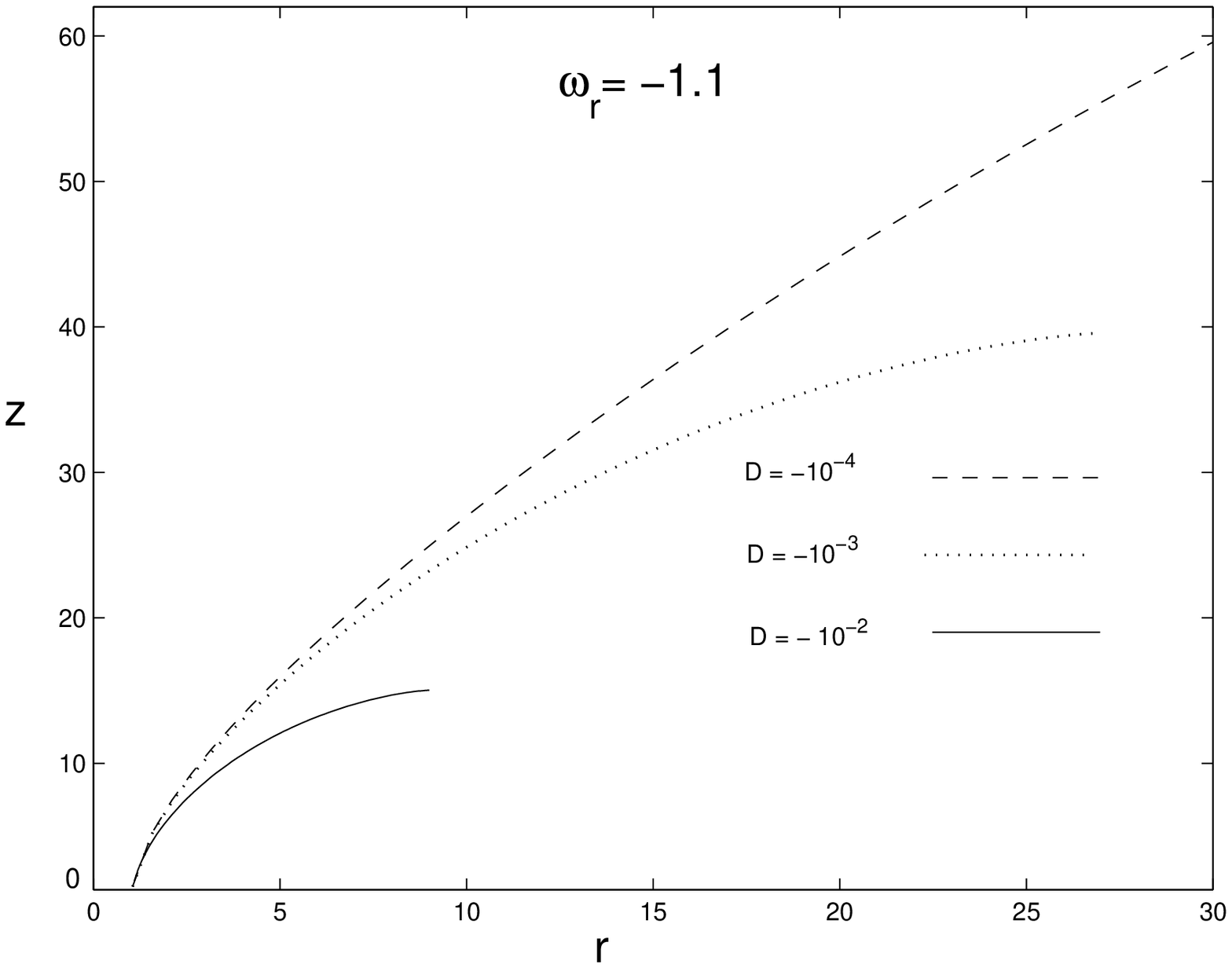}
\caption{\label{WH1} In the figures we show some embedding
diagrams $z(r)$ of two-dimensional sections along the equatorial
plane ($t=$ const, $\theta= \pi/2$) with the help of
Eq.~(\ref{38}) of the traversable evolving wormhole~(\ref{evolving
wormhole final}). For all diagrams the throat is located at
$r_0=1$ and curves are drawn for the specified values of $D$ and
$\omega_r$, taking into account the shape function $b(r)$ of
Eq~(\ref{whevolving}). The range of $r$ is $r_0 <r<r_{max}$, where
$r_{max}$ is given by Eq.~(\ref{rmax}). For a full visualization
of the surfaces the diagrams must be rotated about the vertical
$z$ axis.}
\end{figure}

(v) The shape of a wormhole is determined by $b(r)$ as viewed, for
example, in an embedding diagram in a flat $3$--dimensional
Euclidean space $R^3$. To construct such a diagram of a wormhole,
one considers an equatorial slice ($\theta=\pi/2$) at a fixed
instant of time $t=t_0$ of the geometry. Since the
wormhole~(\ref{evolving wormhole final}) evolves in time, each
such slice will be different for different values of time. In
other words, the shape of the wormhole is determined also by the
scale factor $a(t)$. However, it can be shown that the form of the
wormhole is preserved with time, by using an embedding procedure.
The metric of a such a wormhole slice for $t=t_0=$ const is given
by
\begin{eqnarray}\label{35}
ds^2=a^2(t_0) \left( \frac{dr^2}{1-\frac{b(r)}{r}} +r^2 d\varphi^2
\right),
\end{eqnarray}
where $b(r)$ is given by the first expression of
Eq.~(\ref{whevolving}). One may rewrite this slice by rescaling
the radial coordinate as $\bar{r}=a(t_0) \, r$. Thus the
metric~(\ref{35}) may be rewritten in the following form:
\begin{eqnarray}\label{36}
ds^2=\frac{d\bar{r}^2}{1-\frac{\bar{b}(\bar{r})}{\bar{r}}}
+\bar{r}^2 d\varphi^2,
\end{eqnarray}
where we have introduced the definition $\bar{b}(\bar{r})=a(t_0)\,
b(r)$. Now, we shall embed this slice in a flat $3$--dimensional
Euclidean space $R^3$, which we shall write as
\begin{eqnarray}\label{37}
ds^2=d\bar{z}^2+d\bar{r}^2+\bar{r}^2 d \varphi^2.
\end{eqnarray}
Comparing the metrics~(\ref{36}) and~(\ref{37}) we conclude that
\begin{eqnarray}\label{38}
\frac{d\bar{z}}{d\bar{r}}=\pm \left(\frac{\bar{r}}{\bar{b}}-1
\right)^{-1/2}=\pm \left(\frac{r}{b}-1 \right)^{-1/2}.
\end{eqnarray}
This implies that the evolving wormhole will remain the same size
in the $\bar{z},\bar{r},\varphi$ coordinates.

On the other hand, we also conclude that in order to visualize the
slice $\theta=\pi/2$, $t=t_0$ embedded into the three-dimensional
Euclidean space we must require that the shape function $b(r)$
must be positive and be such that $b(r)/r <1$ in order to
guarantee that the root $\sqrt{r/b(r)-1}$ be real, as for static
wormholes~\cite{Lemos}. In other words we can draw the graph
$\bar{z}=\bar{z}(\bar{r})$ only for $b(r)/r <1$ with $b(r)
>0$. In this case the embedded two dimensional section has a
minimum radius at the throat $r=r_0$ and has the maximum upper
radius at the mouth ($b=0$) of the wormhole. For larger radii
where $b(r) <0$ the embedding process is no longer valid. Notice
that in our case the general metric~(\ref{evolving wormhole}),
with the scale factor and shape function~(\ref{scf})
and~(\ref{whevolving}), is well defined even for $b(r) <0$, being
this spacetime geodesically complete. Thus the requirement
$b(r)>0$ emphases the fact that the importance of the embedding is
near the throat of the wormhole. In our case, as we stated above,
far from the wormhole mouth the space is asymptotically flat. In
principle, if one includes a cosmological constant, the space can
be de-Sitter or anti-de Sitter far from the mouth.

Now in order to maintain the shape of the traversable wormhole the
flaring out condition must be required, i.e.
$d^2\bar{r}/d\bar{z}^2 >0$. So from Eq.~(\ref{38}) we have that
\begin{eqnarray}\label{39}
\frac{d^2\bar{r}}{d\bar{z}^2}=\frac{\bar{b}-\bar{b}^\prime r}{2
\bar{b}^2}=\frac{b-b^\prime r}{2 a(t_0)b^2}>0,
\end{eqnarray}
and taking into account the form of the shape function from
Eq.~(\ref{whevolving}) we obtain
\begin{eqnarray}
\frac{d^2\bar{r}}{d\bar{z}^2}=
-\left(\frac{r}{r_0}\right)^{1/\omega_r} \times
\,\,\,\,\,\,\,\,\,\,\,\,\,\,\,\,\,\,\,\,\,\,\,\,\,\,\,\,\,\,
\,\,\,\,\,\,\,\,\,\, \nonumber \\
\frac{2D\omega_r
r^3(r/r_0)^{1/\omega_r}+(1+\omega_r)r_0(Dr_0^2-1)}{2\omega_r
(Dr^3(r/r_0)^{1/\omega_r}+r_0(1-Dr_0^2))^2},
\end{eqnarray}
which for $D<0$ is always positive, thus satisfying the flaring
out condition for the entire range of the radial coordinate $r$.
So, as we have seen, a distribution of an anisotropic phantom
energy provides the flare-out conditions for the throat of
evolving wormholes.

(vi) Let us now study the range of validity of the radial
coordinate more adequately. From the condition $b(r) \geq 0$,
which we must impose in order to have a good embedding, we obtain
that $b(r)=0$ for
\begin{equation}\label{rmax}
r_{max}=r_0
\left(1-\frac{1}{Dr_0^2}\right)^{\omega_r/(1+3\omega_r)},
\end{equation}
implying that $b(r)\geq 0$ for $r\leq r_{max}$. Thus the wormhole
is located at the range $r_0\leq r \leq r_{max}$, being the throat
at $r_0$. Notice that the radius $r_{max}$ may be made arbitrarily
large by taking $D \rightarrow -0$, but still having an evolving
wormhole.

(vii) In order for this evolving wormhole to be traversable, the
tidal forces experienced by a traveller must not be too great. So
a traveller should feel during its radial journey a tidal
acceleration, between two parts of her body (i.e. head to feet),
which must not exceed by much one Earth gravity. This
traversability criteria was considered in Ref.~\cite{Morris}. In
general the tidal acceleration may be written as (the Greek
indices take the values $0, 1, 2, 3$)
\begin{eqnarray}
\bigtriangleup a^{\hat{\alpha}}=-c^2 R^{\hat{\alpha}}_{\,\,\, \,\,
\hat{\beta} \hat{\gamma} \hat{\delta}} u^{\hat{\beta}}
\xi^{\hat{\gamma}} u^{\hat{\delta}},
\end{eqnarray}
where the vector $\xi^{\hat{\gamma}}$ denotes the separation
between the head and feet of the traveller's body, so
$\xi^{\hat{\gamma}}$ is a spacelike vector.

In order to calculate the tidal acceleration felt by a traveller
we introduce the orthonormal reference frame used by her:
$(e_{\hat{0}^{\prime}},e_{\hat{1}^{\prime}},e_{\hat{2}^{\prime}},e_{\hat{3}^{\prime}})$.
Since in this frame we have that $\xi^{\hat{0}^\prime}=0$ and
$u^{\hat{\beta}^\prime}=\delta^{\hat{\beta}^\prime}_{\hat{0}^\prime}$
for the four velocity, and additionally the Riemann tensor is
antisymmetric in its first two indices, the tidal acceleration is
purely spatial with components (the Latin indices take the values
$1, 2, 3$)
\begin{eqnarray}\label{tidalaccel}
\bigtriangleup a^{\hat{k}^\prime}=-c^2 R^{\hat{k}^\prime}_{\,\,\,
\,\, \hat{0}^\prime \hat{j}^\prime \hat{0}^\prime}
\xi^{\hat{j}^\prime},
\end{eqnarray}
where the spacelike vector $\xi$ may be oriented along any spatial
direction in the traveller's frame.

Now, this traveller moves at a constant speed $v$ with respect to
the observer who uses the orthonormal basis~(\ref{basis quieta})
and who always remains at rest at constant $r$, $\theta$,
$\varphi$. Thus both sets of orthonormal basis vectors are
connected by the standard special relativity Lorentz
transformation as follows~\cite{Morris}:
\begin{eqnarray}\label{LorentzT}
e_{\hat{0}^{\prime}}=\bar{u}= \gamma e_{\hat{t}} \mp \gamma \beta
e_{\hat{r}}, \nonumber \\ e_{\hat{1}^{\prime}}=  \gamma \beta
e_{\hat{r}}\mp \gamma e_{\hat{r}}, \,\,\,\,\,\,\,\,\,\,\,\,
\nonumber \\ e_{\hat{2}^{\prime}}= e_{\hat{\theta}},
e_{\hat{3}^{\prime}}= e_{\hat{\varphi}},
\,\,\,\,\,\,\,\,\,\,\,\,\,\,
\end{eqnarray}
where $\bar{u}$ is the traveller's four velocity,
$\gamma=(1-\beta^2)^{-1/2}$, and $\beta=v/c$. In this case the
vector $e_{\hat{1}}$ points along the direction of travel (towards
increasing radial proper distance $l$).

Thus, from the generic metric~(\ref{evolving wormhole}) (with
$\Phi(t,r)=0$) and the Lorentz transformation~(\ref{LorentzT}),
the relevant Riemann tensor components for~(\ref{tidalaccel}) are
\begin{eqnarray}
R_{{\hat{1}^\prime} \hat{0}^\prime \hat{1}^\prime \hat{0}
}=R_{{(r)} (t) (r) (t)}=\frac{\ddot{a}}{a}, \nonumber \\
R_{{\hat{2}^\prime} \hat{0}^\prime \hat{2}^\prime \hat{0}^\prime}=
R_{{\hat{3}^\prime} \hat{0}^\prime \hat{3}^\prime \hat{0}^\prime}=
\gamma^2 R_{(\theta) (t) (\theta) (t)} \mp \nonumber \\   2
\gamma^2 \beta R_{(\theta) (t) (\theta) (r)}+  \gamma^2 \beta^2
R_{(\theta) (r) (\theta) (r)}= \nonumber \\
\gamma^2 \frac{\ddot{a}}{a}-\frac{\gamma^2 \beta^2}{2 a^2 r^3}
\left( 2 \dot{a}^2 r^3-b+r b^{\prime}\right) .
\end{eqnarray}

If now we consider the size of the traveller's body to be $|\xi|
\sim 2$ (m) and $\mid \bigtriangleup a \mid \leq g_{\oplus}$
($\equiv$ one Earth gravity, i.e. $9,8$ m$/s^2$) the Riemann
tensor components are constrained to be
\begin{eqnarray}\label{R1}
|R_{{\hat{1}^\prime} \hat{0}^\prime \hat{1}^\prime \hat{0}
}|=\left|\frac{\ddot{a}}{a} \right|\leq \frac{g_{\oplus}}{c^2
\times 2 \, m} \simeq \frac{1}{(10^{8} m)^2},
\end{eqnarray}
and
\begin{eqnarray}\label{R2}
|R_{{\hat{2}^\prime} \hat{0}^\prime \hat{2}^\prime
\hat{0}^\prime}|= |R_{{\hat{3}^\prime} \hat{0}^\prime
\hat{3}^\prime \hat{0}^\prime}|=  \nonumber \\ \left |\gamma^2
\frac{\ddot{a}}{a}-\frac{\gamma^2 \beta^2}{2 a^2 r^3} \left( 2
\dot{a}^2 r^3-b+r b^{\prime}\right) \right|\leq \nonumber \\
\frac{g_{\oplus}}{c^2 \times 2 \, m} \simeq \frac{1}{(10^{8}
m)^2}. \nonumber \\
\end{eqnarray}
Notice that, since the wormhole metric evolves with time, the
tidal acceleration also depends on time. In this case the radial
tidal constraint~(\ref{R1}) can be regarded as directly
constraining the acceleration of the expansion of the wormhole,
while the lateral tidal constraint~(\ref{R2}) can be regarded as
constraining the speed $v$ of the traveller while crossing the
wormhole.

\begin{table}[h]
\begin{tabular}{|c|c|c|}
  \hline
  $v_{max}$ & $\omega_{r}$ & $r_0$\\
  \hline
  $ 542$ m/s&  $-1.5$ & $100$ m  \\
\hline
 $1038$ m/s & $-1.1$ & $100$ m  \\
\hline
  $ 1084$ m/s & $-1.5$ & $200$ m   \\
  \hline
$ 2076$ m/s & $-1.1$ & $200$ m   \\
  \hline
\end{tabular}
\caption{This table shows the maximum values of $v_{max}$ at which
the traveller could cross the static wormhole for given values of
$\omega_{r}$ and $r_0$ in order to satisfy the constraint on the
lateral tidal acceleration. \label{tabla15}}
\end{table}

\begin{table}[h]
\begin{tabular}{|c|c|c|c|c|}
  \hline
$t_{min}$ &  $D$ & $v$ & $\omega_{r}$ & $r_0$\\
  \hline
$4.55$ s & $-15$& $50$ m/s&  $-1.1$ & $100$ m  \\
\hline
$1.65$ s & $-0.1$ &$50$ m/s & $-1.1$ & $100$ m  \\
\hline
$4.55 $ s &  $-15 $  & $ 50$ m/s & $-1.1$ & $200$ m   \\
  \hline
$1.65$ s &  $-0.1 $ & $ 50$ m/s & $-1.1$ & $200$ m   \\
  \hline
$6.06$ s &  $-0.1 $ & $ 50$ m/s & $-1.5$ & $100$ m   \\
  \hline
  $4.71$ s &  $-100 $ & $ 50$ m/s & $-1.1$ & $100$ m   \\
  \hline
\end{tabular}
\caption{This table shows the minimum values $t_{mim}$ for the
cosmological time at which the traveller could cross the evolving
wormhole for given values of $D$ ($F=1$), $v$, $\omega_{r}$ and
$r_0$ in order to saturate the constraint on the lateral tidal
acceleration. \label{tabla15A}}
\end{table}

In particular, the evolving wormholes obtained in this paper
evolve with the scale factor~(\ref{scf}). This implies that the
expansion is not accelerated (i.e. $\ddot{a}=0$) and then the
radial tidal acceleration is identically zero, thus satisfying the
constraint~(\ref{R1}). On the other hand, by taking into account
Eqs.~(\ref{evolving wormhole final}) and~(\ref{CRT}) we obtain the
following constraint for the lateral tidal acceleration:
\begin{eqnarray}\label{lateral constraint}
\left|\gamma^2 \beta^2 \frac{\kappa C (1+\omega_r) \,
r^{-(1+3\omega_r)/\omega_r}}{2(\sqrt{-D}t+F)^2}\right| \leq
\frac{g_{\oplus}}{c^2 \times 2 \, m} \simeq \nonumber \\
\frac{1}{(10^{10} cm)^2}.
\end{eqnarray}
It is interesting to note that the lateral tidal acceleration at
fixed $r$ diminishes  with time. Now by taking into account
Eq.~(\ref{rhofinal}) this constraint may be rewritten as
\begin{eqnarray}\label{constraint2}
\left|\frac{1}{2} \, \gamma^2 \beta^2 \kappa (1+\omega_r)\rho
\right|\lesssim \frac{1}{(10^{10} cm)^2},
\end{eqnarray}
thus the lateral tidal constraint~(\ref{lateral constraint}) can
be regarded more exactly as constraining both the speed $v$ of the
traveller and the energy density of the matter threading the
wormhole. By taking into account the expression for the energy
density of Eq.~(\ref{whevolving}) and considering that the motion
of the traveller is nonrelativistic ($v<<c$, $\gamma \approx 1$)
we may rewrite Eq.~(\ref{constraint2}) as follows:
\begin{eqnarray}\label{constraint3}
\left|\frac{v^2 (1+\omega_r)(1-Dr_0^2) }{\omega_r r_0^2
(\sqrt{-D}t+F)^2}\right| \lesssim g_{\oplus}.
\end{eqnarray}
For the static case (i.e. $D=0$ and $F=1$) Eq.~(\ref{constraint3})
gives the following constraint on the speed $v$:
\begin{eqnarray}\label{constraint4}
v \lesssim \sqrt{\frac{g_{\oplus}\, \omega_r \,
r_0^2}{1+\omega_r}}.
\end{eqnarray}
In Table~\ref{tabla15} we show the maximum values of the speed at
which the traveller could cross the static wormhole for some given
values of the $\omega_{r}$ and $r_0$ parameters in order to
satisfy the constraint~(\ref{constraint4}).

In table~\ref{tabla15A} we show for some given values of
$\omega_{r}$, $r_0$, $D$ ($F=1$) and $v$ the minimum values of the
cosmological time $t=t_{min}$ at which it is possible to cross the
evolving wormhole in order to satisfy the
constraint~(\ref{constraint3}) for $t \geq t_{min}$.

(viii) This wormhole solution also may be interpreted as an
interior one~\cite{Arellano}. This implies that one may, in
principle, match the found wormholes to an exterior Kottler
solution (Schwarzschild--de Sitter or Schwarzschild--anti de
Sitter spacetimes) at some matching interface $r_{m}$, where $r_0
< r_m < r_{max}$ (see Fig.~\ref{WH1}), in the spirit of made in
Ref.~\cite{Lemos}, where a procedure is given for matching static
spherically symmetric wormholes to Kottler solution by using
directly the field equations to make the match. This work is in
progress.

\section{Conclusions}
In this paper we have constructed exact evolving wormhole
geometries supported by phantom energy, showing explicitly that
the phantom energy can support the existence of evolving
wormholes. Specifically we have constructed asymptotically flat
evolving wormholes with radial and tangential pressures obeying
barotropic equations of state with constant state parameters. One
interesting feature of these evolving wormholes, supported by an
anisotropic phantom matter, is that they expand with constant
velocity.

\section{Acknowledgements}
This work was supported by CONICYT through Grants FONDECYT N$^0$
1080530 and 1070306 (MC, SdC and PS), and by Direcci\'on de
Investigaci\'on de la Universidad del B\'\i o--B\'\i o (MC and
PL). SdC also was supported by PUCV grant N$^0$ 123.787/2008. P.S.
and J.C. were supported by Universidad de Concepci\'on through
DIUC Grants N$^0$ 208.011.048-1.0 and 205.011.038-1 respectively.


\begin{thebibliography}{2}
\bibitem{Coleman} S. Coleman, Nucl Phys. {\bf 307},
867 (1988).
\bibitem{Strominger} S.B. Giddings and A. Strominger, Nucl. Phys. {\bf B 321},
481 (1988).
\bibitem{Morris} M.S. Morris and K.S. Thorne, Am. J. Phys. {\bf
56}, 395 (1988); M.S. Morris, K.S. Thorne and U. Yurtsever, Phys.
Rev. Lett. {\bf 61}, 1446 (1988).
\bibitem{Novikov} J.D. Novikov, Sov. Phys. JETP {\bf 68}, 439 (1989).
\bibitem{Visser} M. Visser, Lorentzian Wormholes: From Einstein to Hawking, (AIP,
New York, 1995); M. Visser, S. Kar, N. Dadhich Phys. Rev. Lett.
{\bf 90} 201102 (2003); N. Dadhich, S. Kar, S. Mukherjee and M.
Visser, Phys. Rev. D {\bf 65}, 064004 (2002).
\bibitem{Teo} E. Teo, Phys. Rev. D {\bf 58},
024014 (1998); V.M. Khatsymovsky, Phys. Lett. B {\bf 429}, 254
(1998); P. K. F. Kuhfittig, Phys. Rev. D 67, 064015 (2003);
Tonatiuh Matos, D. Nunez Class. Quant. Grav. {\bf 23}, 4485
(2006); Mubasher Jamil, Muneer Ahmad Rashid, Electromagnetic field
around a slowly rotating wormhole arXiv: 0805.0966 [astro-ph].
\bibitem{Kar1} S. Kar, Phys. Rev. D {\bf 49}, 862 (1994).
\bibitem{Kar2} S. Kar and D. Sahdev, Phys. Rev. D {\bf 53}, 722 (1996).
\bibitem{Lobo} F.S.N. Lobo, Exotic solutions in General Relativity: Traversable wormholes
and 'warp drive' spacetimes, e-Print: arXiv:0710.4474 [gr-qc]; A.
V. B. Arellano and F. S. N. Lobo; Class. Quant. Grav. {\bf23},
5811 (2006).
\bibitem{Arellano} A. V. B. Arellano and F. S. N.
Lobo; Class. Quant. Grav. {\bf 23}, 7229 (2006).
\bibitem{Kim} G.P. Perry and R.B. Mann, Gen. Rel. Grav. {\bf 24}, 305 (1992);
S.W. Kim, H.J. Lee, S.K. Kim and J.M. Yang, Phys. Lett. A {\bf
183}, 359 (1993); M.S.R. Delgaty and R.B. Mann, Int. J. Mod. Phys.
D {\bf 4}, 231 (1995); Y.G. Shen and Z.Q. Tan, Annals Phys. {\bf
272}, 1 (1999); W.T. Kim, E.J. Son and M.S. Yoon, Phys. Rev. D
{\bf 70}, 104020 (2004); W.T. Kim, J.J. Oh and M.S. Yoon, Phys.
Rev. D {\bf 70}, 044006 (2004).
\bibitem{Gonzales} P. Gonzales--Diaz, Phys. Lett. B {\bf 247}, 251 (1990);
X. Jianjun and J. Sicong, Mod. Phys. Lett. {\bf 6}, 251 (1990).
\bibitem{Cataldo} Gerard Clement, Gen. Rel. Grav. {\bf 16}, 131, (1984.); M.
Cataldo, P. Salgado and P. Minning, Phys. Rev. D {\bf 66}, 124008
(2002).
\bibitem{Biplab} B. Bhawal and S. Kar, Phys. Rev. D {\bf 46}, 2464 (1992);
G. Dotti, J. Oliva, R. Troncoso, Phys. Rev. D {\bf 76}, 064038
(2007); G. Dotti, J. Oliva, R. Troncoso, Phys. Rev. D {\bf 75},
024002 (2007).
\bibitem{Sayan} S. Kar and D. Sahdev, Phys. Rev. D {\bf 53}, 722 (1996);
A. DeBenedictis and D. Das, Nucl. Phys. B {\bf 653}, 279 (2003).
\bibitem{Barcelo} C.~Barcelo, L.~J.~Garay, P.~F.~Gonzalez-Diaz and G.~A.~Mena
Marugan, Phys.\ Rev.\  D {\bf 53}, 3162 (1996).
\bibitem{Roman15} T.A. Roman, Phys. Rev. D {\bf 47}, 1370 (1993).
\bibitem{Cataldo15} S.~Nojiri, S.~D.~Odintsov and S.~Tsujikawa,
Phys.\ Rev.\  D {\bf 71}, 063004 (2005); P.~F.~Gonzalez-Diaz,
Phys.\ Rev.\  D {\bf 68}, 021303 (2003); P.~F.~Gonzalez-Diaz,
Phys.\ Lett.\  B {\bf 586}, 1 (2004); M.~Cataldo, N.~Cruz and
S.~Lepe, Phys.\ Lett.\  B {\bf 619}, 5 (2005);  G.~Izquierdo and
D.~Pavon, Phys.\ Lett.\  B {\bf 633}, 420 (2006).
\bibitem{Sushkov}S.~V.~Sushkov, Phys.\ Rev.\  D {\bf 71},
043520 (2005)
\bibitem{Lobo1} F.~S.~N.~Lobo, Phys.\ Rev.\  D {\bf 71}, 084011
(2005); F.S.N. Lobo, gr-qc/0603091; P.~K.~F.~Kuhfittig, Class.\
Quant.\ Grav.\  {\bf 23}, 5853 (2006).
\bibitem{Visser1} M. Visser and C. Barcelo, Energy
conditions and their cosmological implications, gr-qc/0001099.
\bibitem{Lemos} J.~P.~S.~Lemos, F.~S.~N.~Lobo and S.~Quinet de Oliveira, Phys.\ Rev.\  D {\bf 68}, 064004
(2003).
\end{thebibliography}
\end{document}